# A New Rb Lamp Exciter Circuit for Rb atomic clocks and Studies on Transition from Ring to Red mode

Savita Singh  Bikash Ghosal* and G M Saxena#

National Physical Laboratory, Dr K.S.Krishnan Road

New Delhi-110012 India

#E-mail: gmsaxena@nplindia.org

## Abstract:

In this paper we describe the development of novel RF exciter circuit for electrode less Rb lamp. The lamp exciter circuit is a RF oscillator with a a new configuration operating at 60 to 65 MHz frequency with 3 to 4 watt power. The Rb lamp is used in exciting the ground state hyperfine transitions in Rb atom in a glass cell placed inside a tuned microwave cavity, As the frequency of these hyperfine transitions is very stable it is used in the development of Rb atomic clock by phase locking the oven controlled crystal oscillator (OCXO) to this atomic transition frequency. The details of the Rb lamp exciter are presented in the paper.The Lamp is ideally operated in ring mode as in this mode the linewidth is narrow and there is no self reversal. However, high temperature and RF excitation power may drive the Rb lamp to red mode which gives rise to line broadening and self reversal. It is the experience that mode change from ring to red deteriorates the atomic signal strength and S/N. In this paper the reasons of mode change are also discussed.

*Space Application Centre

Satellite Road, Ahmedabad, India

## Introduction:

In the Rb atomic clock the Physics Package, where the atomic transitions take place, is the most critical part. The physics package of Rb atomic clock consists of the following components : Rb lamp, Rb absorption cell, microwave cavity, photodiode detector, bifilar heaters with controllers, magnetic field solenoid, magnetic shield, base plate.

## Rb Lamp Exciter:

One of the important components of the Physics Package is Rb lamp. The Rb lamp consists of electrode less Rb bulb and a lamp exciter. The bulb is excited by 80- 100 MHz and 3 to 4 watt RF oscillator which is a Colpitts oscillator run by a D.C. power supply of rating 24 volt and 0.2 to 0.4 current. The coil in the tank circuit of the oscillator is wound inside out for concentrating the RF field in the middle where the Rb bulb is placed. The light intensity and mode of operation of the lamp may be controlled by changing the gain and frequency of the oscillator with the help of a resistor $R_1$, capacitors $C_2$ and $C_4$ respectively. It has been observed that on increasing the temperature of the Rb bulb beyond 140 $^0$C the mode is changed to Red and the frequency of the exciter is reduced due to the excessive presence of the Rb atoms in the bulb. The cold sealing notch is not capable of retaining bulk of the the Rb atoms due to high velocity of Rb atoms. This fluctuating frequency of the oscillator due to excessive random motion of the Rb atoms leads to intensity fluctuations. So it is very important to operate the Rb bulb around 110 $^0$C

### Components Used:

- Three resistors of values: $R_1$ =5 Ω; $R_2$ =2.2 kΩ; $R_3$ =2 kΩ (1/4 watt each)
- Three inductors of values: $L_1$ =0.43 μH; $L_2$ =4.7μH; $L_3$ =4.7 μH
- Three capacitors of values: $C_1$ =2 nf ; $C_2$= 2.13 pf ( 3.9pf in parallel 4.7pf); Coupling capacitor: $C_c$ =100 pf
- Rf Power transistor:2N3375 NPN
- Supply voltage: 15 volt- 24 volt
- RF coil of value: 1.45 μH
- Feedback capacitor

# Working Principle:

The discharge lamp drives the optical pumping process, Rb discharge lamp emits light at frequencies that connect Rb's excited states with each ground state hyperfine level. Due to a coincidence of nature, spectral lines associated with Rb's F=2 state

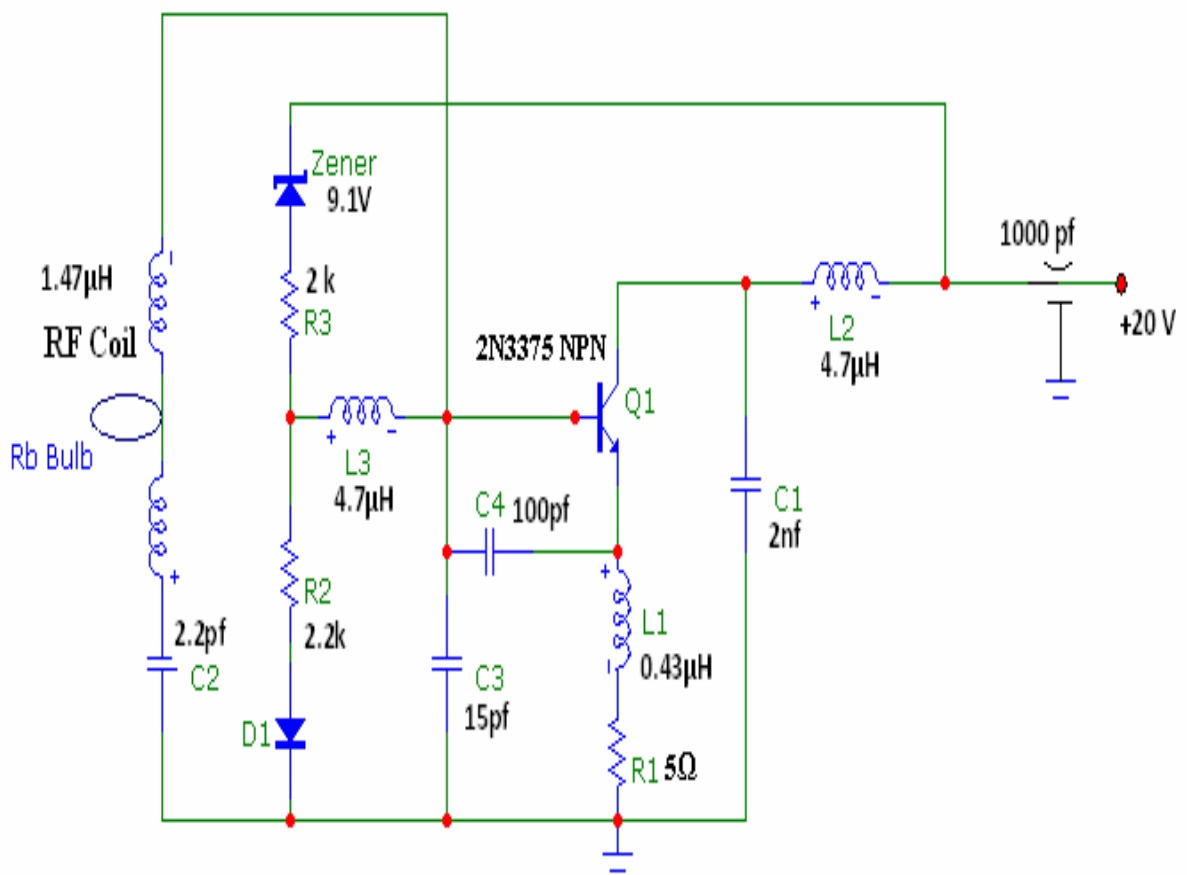

Lamp Exciter Ckt Diagram

In the Rb atomic clock with the conventional gas discharge lamps the performance of the clock is highly sensitive to the behaviour of the Rb lamp exciter. We have developed a circuit which is a variant of the Colpitt's oscillator. In the Colpitt's oscillator the capacitors are split and the frequency of the oscillator depends on the parallel combination of the capacitor. The Rb lamp exciter is shown in the figure-1

The frequency of a Colpitt's oscillator is given by

$$f = \pi\sqrt{(C_2+C_3)}/[\sqrt{LC_2C_3}] \qquad (1)$$

The symbols are same as shown in fig-1. In the Colpitt's oscillator circuit, the capacitor $C_4$ is not present. The circuit given in figure-1 has been designed based on the observation that with the Colpitt's oscillator circuit, generally, the excitation of the Rb lamp needs more power and higher oscillator frequency (90 to 120MHz). However, with the inclusion of the capacitor $C_4$ in the circuit, the Rb bulb excites with smaller power and at lower frequency (55 to 70 MHz). It is also observed that even when $C_3$ is removed the circuit works and Rb atoms can be excited without any problem. The circuit without $C_3$ is hybrid of Hartley and Colpitt's (H-C hybrid) oscillators. This is a circuit with new configuration and delivers more rf power at lower D.C input power. In the space applications where the Rb atomic clocks are used, such a hybrid Hartley Colpitt Rb lamp exciter circuit is very useful for its ease of operation and low power requirements. The frequency of H-C hybrid oscillator circuit is given by the eqn.1 with $L=L_{rfcoil}$.

## Rb Lamp Modes and Discussion on Mode Changes

The Rb lamp used in Rb atomic clocks are maintained at appropriate temperature for obtaining the desired narrow emission lines for optical pumping the Rb atoms. The temperature plays important role in determining emission line profile. At the nominal temperature (110-120 $^o$C) the Lamp operates in the ring mode. In this mode the central part of the lamp appears to be whitish surrounded by a purple ring of excited Rb atoms. This is an ideal mode and the line is very narrow. As the temperature is increased there is change from the ring to the red mode. In the red mode, the whole of the lamp appears purple with the line being broadened and self-reversed. This is an undesired mode. We shall discuss the reasons of the transition from the ring to the red mode. The phenomenon has been studied by various groups. Shah [1] and Camparo [2] have described the mode transition due to the radiation trapping. Their explanations are not fully satisfactory. Camparo has made ad hoc assumption that P-levels of Rb atoms become metastable. And it gives rise to radiation trapping.

It appears that to describe the mode transition we should consider the coherent population trapping (CPT) of Rb atoms in the ground state hyperfine levels. This is possible because as Dicke [3] described in his seminal work on the presence of coherence in the spontaneous radiation process. The coherence in the spontaneous emission leads to confinement of unexcited Rb atoms to the ground state (g.s) hyperfine sub-levels due to the destructive interference between the two radiation lines originating from the transitions from ground state hyperfine levels $^5S$ F=2 and F=1 to $^5P_{1/2}$ in the Rb lamp. This trapping of some of the atoms in the g.s hyperfine levels is equivalent to the situation as if there is radiation trapping in the excited state as inferred by Camparo. However, Camparo has made a assumption that the excited state behaves as metastable state. While this reduced intensity of the Rb lamp is the result of the trapping of the atoms in the ground state hyperfine levels due to the partial coherence in the spontaneous emission. The population trapping finally results in the loss of optical pumping efficiency in the Rb absorption cell used in the Rb atomic clocks. The detailed description of the phenomenon responsible for the mode change and self reversal will be described in an other paper.